\documentclass[twocolumn,aps,showpacs,amsmath,amssymb]{revtex4}

\usepackage{graphicx}
\usepackage{dcolumn}
\usepackage{bm}

\begin{document}

Journal Ref.: V. Yu. Denisov, Phys. Rev. C 88, 044608 (2013).

\title{
The multidimensional model of cluster radioactivity
}%

\author{
V. Yu. Denisov
}%

\affiliation{%
Institute for Nuclear Research, Prospect Nauki 47,
03680 Kiev, Ukraine
}%

\begin{abstract}
The cluster decays $^{228}$Th $\rightarrow \, ^{208}$Pb + $^{20}$O, $^{232}$U $\rightarrow \, ^{208}$Pb + $^{24}$Ne, $^{236}$Pu $\rightarrow \, ^{208}$Pb + $^{28}$Mg, $^{242}$Cm $\rightarrow \, ^{208}$Pb + $^{34}$Si are considered in the framework of the multidimensional cluster preformation model. The macroscopic potential energy surface related to the interaction between the cluster and the residue nucleus is evaluated in the framework of the nonlocal $\hbar^4$ extended Thomas-Fermi approach with Skyrme and Coulomb forces. The shell-correction to the macroscopic potential energy is also taken into account. The dynamical surface deformations of both the cluster and the residue nucleus are taken into consideration at the barrier penetration path. The heights of saddle points related to deformed nuclear shapes are lower than the barrier height between the spherical cluster and residue nuclei; therefore the dynamical deformations of nuclei increase the barrier penetrability and reduce the half-life of cluster decay. The shell correction contribution into the potential energy between cluster and residue nucleus is important for both the potential landscape and the half-life evaluation. The experimental values of cluster decay half-lives are well reproduced in the model.
\end{abstract}


\maketitle

\section{Introduction}
Cluster decay of nuclei was predicted by Sandulesku, Poenaru, and Greiner in 1980 \cite{spg} and observed in experiments four years later \cite{rj,abgnnos,ghhssv}. Since then, a lot of various experimental and theoretical works have been done; see Refs. [5-61] and papers cited therein. Reviews of diverse aspects of cluster emission from heavy nuclei can be found in Refs. \cite{zmtfkc,t,bg,hrm,kkt,pg,trm,delion,del13,li13,xpw13,pgg13}.
The detailed experimental information on emission of various clusters from different nuclei has been accumulated in Refs. \cite{zmtfkc,t,trm,bg}.

Several various approaches have been proposed for the description of the cluster decay of nuclei. Cluster emission from heavy nuclei is considered as a very asymmetric fission process in Refs. \cite{spg,pp,is,psgmg,rgd,rm,dtgdgrg,hrm,pg,delion,zdrzl,msd,snr}. Cluster decay is treated in the framework of the cluster-preformation model in Refs. \cite{ss,ld,bw,bm,bmp,bmps,basu,rxw,dd,sspu,ssbb,sbpg,delion,rnb,kaa,raggdopr,stkort,nr,sbs,ssg,agkss,k,ipwbm,xpw13,zzyq}, when clusters exist in nuclei and emission of one is similar to alpha-particle emission in alpha-decay of heavy nuclei \cite{delion,dk}. There are diverse microscopic descriptions of cluster emission, too \cite{llivd,kkt,delion,wr}. The half-life of cluster decay can be also estimated by using various empirical relations \cite{h,pg,trm,nrdx,qxlwza,rxw,delion,pgg,spr}.

Note that cluster emission and heavy-ion fusion reactions are mutually inverse processes. Therefore, it is possible to use such mutually inverse processes for better definition of the potential between the cluster and the residue nucleus \cite{stkort}. Similar mutually inverse processes have been used for accurate evaluation the alpha-nucleus potential too \cite{dk}. Both data sets for cluster emission and for elastic scattering are applied to improve the accuracy of the potential \cite{raggdopr}.

Subbarrier heavy-ion fusion cross sections are strongly enhanced by coupling to the surface vibrational states as well as to the ground-state surface deformations of fusing nuclei [64-77]. For example, the quadrupole and octupole surface vibrations in $^{208}$Pb enhance strongly the subbarrier heavy-ion fusion of $^{208}$Pb and $^{16}$O \cite{dhmhlttn,yhr}. Note that $^{208}$Pb or neighboring nuclei are typical residue nuclei of cluster-emission processes. Therefore, it is reasonable to consider the influence of surface deformations of both the cluster and the residue nucleus on barrier penetration at cluster emission. Such influence is discussed in the framework of a simple schematic approach in Ref. \cite{ld}. Recently, the effect of the ground state deformation of both the cluster and the residue nucleus on the cluster emission half-life has been studied in Refs. \cite{agkss,ssbb,ssg,k}.

The shape deformations of both the cluster and the residue nucleus can be changed during barrier penetration at the cluster emission process. The effect of such dynamical deformation before the scission has been considered in Ref. \cite{msd}, but both fragments are spherical after scission in this model. Therefore, it is interesting to consider the dynamical deformations of both the cluster and the residue nucleus along the barrier penetration trajectory both before and after the scission (or touching point of two nuclei). A similar task is considered in the framework of the multidimensional fusion model \cite{d91,dr,dr96} to describe the subbarrier fusion of heavy ions. The subbarrier fusion cross sections for various heavy-ion reactions are well described in the framework of the multidimensional fusion model \cite{d91,dr,dr96}. Therefore it is reasonable to apply the basic ideas of multidimensional fusion model to describe cluster emission.

The total energy of the residue nucleus and the cluster is evaluated in the framework of the macroscopic-microscopic approach in our approach.

We determine the macroscopic energies of the residue nucleus and the cluster at various surface deformations before and after the scission in the framework of the nonlocal extended Thomas-Fermi (ETF) approximation with all $\hbar^4$ correction terms with the Skyrme and Coulomb forces \cite{bgh}. The density distributions of protons and neutrons are defined by a new parametrization, which describes two deformed separated nuclei as well as one deformed nucleus with a cluster ready for emission.

The microscopic part of the cluster-nucleus interaction related to shell-corrections \cite{s} is added to the macroscopic one for evaluation of the total potential energy of the system. The shell-corrections related to both the single-particle-spectrum non-uniformity around the Fermi energy and the pairing corrections \cite{s} are added to the macroscopic part of interaction energy between the cluster and the residue nucleus. The nonuniformity of single particle spectra near the Fermi-surface is very important for evaluation of the binding energy of nuclei, deformation energy, fission trajectory, and the determination of the magic numbers \cite{s,apdt,skmstar,kp}.

Such approximation for total energy of a nuclear system is successfully used for evaluation of the macroscopic-microscopic atomic mass table \cite{apdt} and fission barrier properties \cite{skmstar}. Therefore, it is reasonable to apply this accurate approach to describe the cluster emission process.

We consider decays related to emission of even-even cluster nuclei and the residue nucleus $^{208}$Pb with well-known experimental values of the cluster decay half-lives: $^{228}$Th $\rightarrow \, ^{208}$Pb + $^{20}$O, $^{232}$U $\rightarrow \, ^{208}$Pb + $^{24}$Ne, $^{236}$Pu $\rightarrow \, ^{208}$Pb + $^{28}$Mg, $^{242}$Cm $\rightarrow \, ^{208}$Pb + $^{34}$Si.

The axial symmetry of distribution of the density during cluster emission is proposed, because the axial symmetry of the density distribution accelerates the numerical calculations. This proposal is exact for the cases $^{228}$Th $\rightarrow \, ^{208}$Pb + $^{20}$O and $^{242}$Cm $\rightarrow \, ^{208}$Pb + $^{34}$Si, because the ground-state shapes of nuclei $^{208}$Pb, $^{20}$O and $^{34}$Si are spherical or nearly spherical \cite{mnms}. The dynamic of density distribution at cluster decay of $^{232}$U related to the oblate nucleus $^{24}$Ne is axial symmetric, too, because such orientation of spherical and oblate nuclei leads to the lowest value of the Coulomb interaction energy at large distances \cite{d90}. The effect related to a nonaxial density distribution may take effect at the cluster decay due to rotation of cluster nucleus with a deformed ground-state shape during barrier penetration; however, we ignore this effect here for the sake of simplicity.

The multidimensional model for cluster decay is presented in the next section. Section III is related to discussion of results and the conclusions are given in Sec. IV.

\section{The multidimensional model}

\subsection{The macroscopic interaction potential between nuclei}

The macroscopic part of interaction potential energy $V_{\rm macro}(R,\xi_1,\xi_2)$ between deformed residue and cluster nuclei is
\begin{eqnarray}
V_{\rm macro}(R,\xi_1,\xi_2) = E(R,\xi_1,\xi_2) - E_1 - E_2 ,
\end{eqnarray}
where $E_1$, $E_2$ are the binding energies of the noninteracting residue and cluster nuclei, respectively, $E(R,\xi_1,\xi_2)$ is the energy of interacting nuclei a distance $R$ between the mass centers of the separated nuclei, and $\xi_1$ and $\xi_2$ are the deformation parameters of the residue and cluster nuclei, which will be specified later.

We need a simple and accurate approach for obtaining $V_{\rm macro}(R,\xi_1,\xi_2)$ for description of cluster emission. Therefore we evaluate the potential energy of two nuclei in the framework of the semiclassical energy-density approximation, which includes the Skyrme and Coulomb interactions as well as the kinetic energies of protons and neutrons obtained in the ETF approach \cite{bgh}. A similar approximation was successfully used for evaluation of the atomic masses \cite{apdt}, the fission barrier characteristics \cite{bgh}, and the nucleus-nucleus potentials \cite{d2002,dn,dnest1}. The barrier heights of the nucleus-nucleus potentials for various systems evaluated in the framework of such approximation well agree with the empirical ones \cite{d_barr}.

The binding energies in Eq. (1) are determined by the energy density functional ${\cal E}[ \rho_{p}({\bf r}), \rho_{n}({\bf r})]$, i.e.,
\begin{eqnarray}
E(R,\xi_1,\xi_2) = \int {\cal E}[\rho_{p}(\mathbf{r},R,\xi_1,\xi_2), \rho_{n}(\mathbf{r},R,\xi_1,\xi_2)] d{\bf r}, \\
E_1 = \int \; {\cal E}[ \rho_{1p}({\bf r}), \rho_{1n}({\bf r})] \;
d{\bf r}, \\
E_2 = \int \; {\cal E}[ \rho_{2p}({\bf r}), \rho_{2n}({\bf r})] \;
d{\bf r},
\end{eqnarray}
where $\rho_{1p}({\bf r})$, $\rho_{2p}({\bf r})$, $\rho_{1n}({\bf r})$ and $\rho_{2n}({\bf r})$ are the proton and neutron densities of the non-interacting residue and cluster nuclei, while $\rho_{p}(\mathbf{r},R,\xi_1,\xi_2)$ and $\rho_{n}(\mathbf{r},R,\xi_1,\xi_2)$ are the proton and neutron densities of the interacting nuclei.

\subsection{Energy-density functional}

According to Ref. \cite{bgh}, the following expression for the energy-density functional has been deduced
\begin{eqnarray}
{\cal E}[ \rho_{p}({\bf r}), \rho_{n}({\bf r})] =
\frac{\hbar^2}{2m} [\tau_p({\bf r}) +\tau_n({\bf r})] + {\cal
V}_{\rm Sk}({\bf r}) + {\cal V}_{\rm C}({\bf r}).
\end{eqnarray}
The kinetic parts for protons ($i=p$) and
neutrons ($i=n$) are given by
\begin{eqnarray}
\tau_{i}({\bf r}) = \tau_{i TF}({\bf r}) + \tau_{i 2}({\bf r}) + \tau_{i 4}({\bf r}),
\end{eqnarray}
where $\tau_{i TF}({\bf r})$ is the Thomas-Fermi contribution to the kinetic-energy density functional and $\tau_{i 2}({\bf r})$ and $\tau_{i 4}({\bf r})$ are semiclassical $\hbar^2$ and $\hbar^4$ correction terms to the kinetic-energy-density functional for the nonlocal case, respectively. The nuclear interaction part ${\cal V}_{\rm Sk}({\bf r})$ results from the Skyrme force and reads
\begin{eqnarray}
{\cal V}_{\rm Sk}({\bf r}) \; = \; \frac{t_0}{2} \;
[(1+\frac{1}{2}x_0) \rho^2 -
(x_0+\frac{1}{2}) (\rho_p^2+\rho_n^2)] \;\\
+\frac{1}{12} t_3 \rho^\alpha [(1+\frac{1}{2}x_3 )\rho^2 -
(x_3+\frac{1}{2}) (\rho_p^2+\rho_n^2) ] \nonumber \\
+\frac{1}{4} [t_1(1+\frac{1}{2}x_1)+t_2(1+\frac{1}{2}x_2)] \tau \rho
\nonumber \\
+\frac{1}{4} [t_2(x_2+\frac{1}{2}) - t_1(x_1+\frac{1}{2})] (\tau_p
\rho_p+\tau_n \rho_n)
\nonumber \\
+\frac{1}{16}[3t_1(1+\frac{1}{2} x_1)-t_2(1+\frac{1}{2}x_2)] (\nabla
\rho)^2 \nonumber \\
- \frac{1}{16}[3t_1(x_1+\frac{1}{2} )+t_2(x_2+\frac{1}{2})] (\nabla
\rho_n)^2 +(\nabla \rho_p)^2 ) \nonumber \\
- \frac{W_0}{2} \left[ \mathbf{J} \cdot \mathbf{\nabla}\rho
+ \mathbf{J}_n \cdot \mathbf{\nabla}\rho_n
+ \mathbf{J}_p \cdot \mathbf{\nabla}\rho_p \right] , \nonumber
\end{eqnarray}
where $t_0$, $t_1$, $t_2$, $x_0$, $x_1$, $x_2$, $\alpha$ and $W_0$
are the Skyrme-force parameters, $\mathbf{J}_i$ are the spin-orbit densities, $\rho=\rho_p+\rho_n$, $\tau=\tau_p+\tau_n$ and $\mathbf{J}= \mathbf{J}_p + \mathbf{J}_n$. The Coulomb-energy density is
determined by
\begin{eqnarray}
{\cal V}_{\rm C}({\bf r}) = \frac{e^2 }{2} \rho_p({\bf r}) \int \;
\frac{\rho_p ({\bf r}' )}{|{\bf r}-{\bf r}' |}
d {\bf r}' \\
-\frac{3e^2}{4} \left( \frac{3}{\pi} \right)^{1/3} (\rho_p({\bf
r}))^{4/3}, \nonumber
\end{eqnarray}
where the last term is the local approximation to the exchange
contribution, and $e$ is the proton charge.

\subsection{Parametrization of density distribution}

It is difficult to find the density distributions of protons and neutrons by solving the integro-differential variational Lagrange equations in the framework of the nonlocal $\hbar^2$ ETF approach for the case of a spherical nucleus \cite{dnest2}. The neutron and proton densities in nuclei in the framework of the nonlocal $\hbar^4$ ETF approach are found by using trial functions only \cite{bgh}. The proton and neutron density distributions for system of interacting nuclei have not yet been evaluated in the framework of the ETF. As a rule, the densities of interacting nuclei are parametrized according to specific physical conditions of the reaction.

The sudden (frozen-density) approximation for density distributions of interacting nuclei is often used at evaluation of the nucleus-nucleus potentials in the framework of energy density \cite{d2002,dn,dnest1} and double-folding \cite{fl,folding} approaches. The sudden approximation is applied to the fast nucleus-nucleus collisions, when the proton and neutron densities cannot quickly relax \cite{dn}. The proton or neutron densities at the fixed point of space are the sum of the corresponding nucleon densities of each nucleus at this point for the case of the sudden approximation. As the result, the nucleon density can exceed the equilibrium density of nuclear matter $\overline{\rho}$ in some space region at small distances between nuclei.

The cluster decay is a deep-subbarrier process; therefore, it is very slow. During this process the proton and neutron densities are relaxed and the densities of nuclei cannot be simply presented as the sum of nucleon densities of interacting nuclei. The relaxed density distributions should satisfy the following conditions:
\begin{itemize}
\item[--] The values of density in any point of space cannot exceed the equilibrium density of nuclear matter, because the compressibility modulus of nuclear matter strongly prevents excess of $\overline{\rho}$.
\item[--] The values of relaxed density at any point should be smaller than the one at the sudden approximation, but larger than the density values of any of the interacting nuclei at this point.
\end{itemize}
Taking into account these conditions we parametrize the proton (neutron) density of the interacting nuclei at point $\mathbf{r}$ as
\begin{eqnarray}
\rho_{p(n)}(\mathbf{r},R,\xi_1,\xi_2) =
\rho_{1p(n)}(\mathbf{r},\xi_1)
+\rho_{2p(n)}(\mathbf{r},R,\xi_2) \\
- \frac{2 \; \rho_{1p(n)}(\mathbf{r},\xi_1) \;
\rho_{2p(n)}(\mathbf{r},R,\xi_2)}
{\rho_{1p(n)} +\rho_{2p(n)}}. \nonumber
\end{eqnarray}
Here $\rho_{1p(n)}(\mathbf{r},\xi_1)=\rho_{1p(n)} f_{1p(n)}(\mathbf{r},\xi_1)$ and $\rho_{2p(n)}(\mathbf{r},R,\xi_2)=\rho_{2p(n)} f_{2p(n)}(\mathbf{r},R,\xi_2)$ are the densities of the proton (neutron) of the residue and cluster nuclei, respectively.

Let us consider the two opposite limits of parametrization (9) in detail.

In the case of well overlapped nuclei at the point inside the nuclei, where \begin{eqnarray}
\rho_{1p(n)}(\mathbf{r},\xi_1) \approx \rho_{1p(n)} = \rho_{p(n)}(1-\eta_1), \nonumber \\
\rho_{2p(n)}(\mathbf{r},R,\xi_2)\approx \rho_{2p(n)} = \rho_{p(n)}(1-\eta_2), \nonumber \\
f_{1}(\mathbf{r},\xi_1) \approx f_{2}(\mathbf{r},R,\xi_2) \approx 1 , \nonumber \\ 0<\eta_{1(2)}<<1, \nonumber
\end{eqnarray}
we get $\rho(\mathbf{r},R,\xi_1,\xi_2) \approx \rho_{p(n)}[1-(\eta_1+\eta_2)/2] < \rho_{p(n)}$. Here, $\rho_{p(n)}$ is the proton (neutron) density in the center of parent nucleus. The saturation conditions of the proton and neutron densities related to Eq. (9) and densities of cluster and residue nucleus are fulfilled, because these saturation conditions in parent nuclei are fulfilled initially.

Note, that the value of total density at $r=0$ in the case of strongly overlapped nuclei $R=0$ is close to the double density of nuclear matter $2 \overline{\rho}$ in the frozen-density approximation. In contrast to this, the density obtained by using Eq. (9) is $\rho(\mathbf{r}=0,R,\xi_1,\xi_2) \approx (\rho_p+\rho_n)[1-(\eta_1+\eta_2)/2] < \rho_p+\rho_n < \overline{\rho}$. Therefore, the saturation condition of the total density is fulfilled in our approach.

In the opposite case of well separated nuclei in the point between them, where \begin{eqnarray}
\rho_{1}(\mathbf{r})\approx \overline{\rho}\eta_1, \nonumber \\ \rho_{2}(\mathbf{r},R,\xi_2)\approx \overline{\rho}\eta_2,\nonumber \\
\rho_{1} \approx \rho_{2} \approx \overline{\rho}, \nonumber
\end{eqnarray}
we find that
\begin{eqnarray}
\overline{\rho}\eta_{1(2)} < \rho(\mathbf{r},R,\xi_1,\xi_2) = \overline{\rho}(\eta_1+\eta_2 - \eta_1 \eta_2) < \nonumber \\ < \overline{\rho}\eta_1 +\overline{\rho}\eta_2 \approx \rho_{1}(\mathbf{r}) + \rho_{2}(\mathbf{r},R,\xi_2).
\nonumber
\end{eqnarray}
So, parametrization (9) satisfies the proposed conditions.

Prescription (9) drastically simplifies the numerical calculations of the relaxed potential energy surface $V_{\rm macro}(R,\xi_1,\xi_2) $ in the framework of the ETF approach with the Skyrme and Coulomb forces.

The proton (neutron) density distribution of the residue nucleus is
\begin{eqnarray}
f_{1p(n)}(\mathbf{r},\xi_1) = \\ 1/\left[1+\exp\left(dist(\mathbf{r},\xi_1,S_{1p(n)})/d_{1p(n)}\right)\right],
\nonumber
\end{eqnarray}
where $d_{1p(n)}$ are the diffuseness parameters, $dist(\mathbf{r},\xi_1,S_{1p(n)})$ is the distance between the point $\mathbf{r}$ and the proton (neutron) surface of the residue nucleus ($S_{1p(n)}$), which we describe by the axial ellipsoid
\begin{eqnarray}
\left[\frac{\varrho}{R_{1p(n)}(1-\xi_1)} \right]^2+ \left[\frac{z}{R_{1p(n)}(1+2 \xi_1)} \right]^2 = 1.
\end{eqnarray}
Here, $\varrho$ and $z$ are the cylindrical coordinates, $R_1$ is the radius parameter, and $\xi_1$ is the deformation parameter, which is proportional to the widely used the quadrupole deformation parameter $\beta$, which is related to the spherical harmonic function $Y_{20}(\vartheta)$ ($\xi \approx \sqrt{\frac{5}{16 \pi}} \beta$). Note that the Fermi distribution (10) fits well the proton and neutron Hartree-Fock densities in nuclei \cite{bgh} and the experimental charge densities in various nuclei \cite{bgh,vjv}.

The density distribution of the cluster nucleus is
\begin{eqnarray}
f_{2p(n)}(\mathbf{r},R,\xi_2) = \\ 1/\left[1+\exp\left(dist(\mathbf{r},R,\xi_2,S_{2p(n)})/d_{2p(n)}\right)\right], \nonumber
\end{eqnarray}
where $d_{2p(n)}$ are the diffuseness parameters and $dist(\mathbf{r},R,\xi_2,S_{2p(n)})$ is the distance between the point $\mathbf{r}$ and the proton (neutron) surface of the cluster ($S_{2p(n)}$), which we also describe by the axial ellipsoid
\begin{eqnarray}
\left[\frac{\varrho}{R_{2p(n)}(1-\xi_2)} \right]^2+ \left[\frac{z-R}{R_{2p(n)}(1+2 \xi_2)} \right]^2 = 1.
\end{eqnarray}

For the sake of simplicity, we consider the same values of deformation parameters for the proton and neutron subsystems for the same nucleus.

We can easily find the parameters $\rho_{1(2)p(n)}, R_{1(2)p(n)}, d_{1(2)p(n)}$ by minimizing the binding energies $E_1$ and $E_2$ for non-interacting spherical nuclei 1 and 2 at fixed values of protons $Z_1, Z_2$ and neutrons $N_1, N_2$ in these nuclei; see also Ref. \cite{bgh}.

We use the same parameters values for central densities $\rho_{1(2)p(n)}$ and diffuseness $d_{1(2)p(n)}$ as the ones for non-interacting spherical nuclei at any values $R$, $\xi_1$ and $\xi_2$.

The conservation condition of proton $Z_2$ (neutron $N_2$) number in the cluster
\begin{eqnarray}
\int d \mathbf{r}\; \rho_{2p(n)} f_{2p(n)}(\mathbf{r},R,\xi_2) = Z_2(N_2)
\end{eqnarray}
fixes the value of the radius $R_{2p}$ ($R_{2n}$) for a deformed cluster at any value of $R$ and $\xi_2$, respectively. Knowing $\rho_{1(2)p(n)}$, $d_{1(2)p(n)}$, and $R_{2p(n)}$ we can find values $R_{1p(n)}$ from the conservation conditions of the total numbers of protons $Z_1+Z_2$ and neutrons $N_1+N_2$ in the system of interaction nuclei, which are
\begin{eqnarray}
\int d \mathbf{r}\; \rho_{p(n)}(\mathbf{r},R,\xi_1,\xi_2) = Z_1+Z_2 \; (=N_1+N_2).
\end{eqnarray}
These conditions take into account that the densities of nuclei at small values of $R$ are distributed in space according to ansatz (9).

The density distribution of the cluster nucleus is determined in space by $\rho_{2p(n)}$, $d_{2p(n)}$, $\xi_2$ and conditions (14) at any $R$ in our approach. In contrast to this, the density distribution of the residue nucleus depends on $R$, $\rho_{1p(n)}$, $d_{1p(n)}$, $\xi_1$ and $\rho_{2p(n)}$, $d_{2p(n)}$, $\xi_2$. This agrees with the cluster preformation model, when ready for emission the cluster exists in the parent nucleus.

By using ansatz (9) for the relaxed density parametrization, shapes (11) and (13), conditions (14) and (15) we can describe densities of deformed residue nucleus and cluster at any values $R$, $\xi_1$, and $\xi_2$.

The neck is often described by using an additional parameter in various cluster emission or fission models \cite{pg,msd,kp}. In our approach we do not introduce additional parameter for the neck, which is smoothly described due to diffuse distribution of densities in both nuclei (10), (12) and relaxed density ansatz (9).

Substituting Eqs. (9), (10) and (12) into Eqs. (5)--(8) we can easily evaluate the energy density at any values of collective coordinates $R,\xi_1,\xi_2$ and, therefore, the macroscopic interaction potential energy between the cluster and the residue nuclei $V_{\rm macro}(R,\xi_1,\xi_2)$ with the help of Eqs. (1)--(4).

The value of the binding energy of $^{208}$Pb obtained in our energy density approach with Fermi distributions of proton and neutron densities and the SkM$^{\star}$ parameter set of Skyrme force \cite{skmstar} equals 1603.6 MeV; see also \cite{bgh}. For the sake of checking the accuracy of relaxed density prescription (9) and density saturation properties we evaluate the binding energy of $^{208}$Pb in the cluster representation related to decay $^{208}$Pb $\rightarrow \; ^{174}$Er $+ \;^{34}$Si. Residue nucleus $^{174}$Er and cluster $^{34}$Si at $R=0$ and $\xi_1=\xi_2=0$ compose the spherical parent nucleus $^{208}$Pb. The binding energy of $^{208}$Pb obtained with the help of the cluster representation, SkM$^{\star}$ parameter set of the Skyrme force, and Eqs. (2), (5)--(15) equals 1601.2 MeV. The difference 2.4 MeV between two values of binding energy of $^{208}$Pb is very small in comparison to the value of the binding energy. (This difference is induced by slightly different values of the parameters $\rho_{1(2)p(n)}, R_{1(2)p(n)}, d_{1(2)p(n)}$ found by minimization of the binding energies of $^{208}$Pb, $^{174}$Er and $^{34}$Si.) The saturation properties of nucleon densities fulfill, when $^{208}$Pb is described as a single isolated nucleus by definition. Due to a small difference of binding energies of $^{208}$Pb evaluated in single-nucleus and cluster representations, the saturation properties for proton and neutron densities are satisfied. So, our prescription for relaxed density of two nuclei (9) is sufficiently accurate.

\subsection{The total interaction potential between nuclei and microscopic corrections}

Applying Strutinsky's shell-correction prescription \cite{s} to the system of interacting nuclei, we get the total interaction potential energy in the form (see also \cite{dr96})
\begin{eqnarray}
V_{\rm tot}(R,\xi_1,\xi_2) &=& V_{\rm macro}(R,\xi_1,\xi_2) + V_{\rm micro}(R,\xi_1,\xi_2) \nonumber \\ &=& E_{12}(R,\xi_1,\xi_2) - E_1 - E_2 \nonumber \\ &+& \delta E_{12}(R,\xi_1,\xi_2) - \delta E_1 - \delta E_2.
\end{eqnarray}
Here $\delta E_{12}(R,\xi_1,\xi_2)$ is the shell-correction for system of interacting nuclei, $ \delta E_1$ and $\delta E_2$ are the ground-state shell-corrections of non-interacting residue nucleus and cluster, respectively. Shell-corrections $\delta E_{12}(R,\xi_1,\xi_2)$, $ \delta E_1$, and $\delta E_2$ include the proton and neutron shell-corrections related to both the single-particle-spectrum nonuniformity around the Fermi energy and the pairing corrections \cite{s}.

It is obvious that mutual influence of nuclei on their single-particle spectra is negligible at large distances between nuclei, therefore
\begin{eqnarray}
\delta E_{12}(R,\xi_1,\xi_2)|_{R \rightarrow \infty} = \delta E_1(\xi_1) + \delta E_2 (\xi_2).
\end{eqnarray}
Here $\delta E_1(\xi_1)$ and $\delta E_2 (\xi_2)$ are the shell corrections of residue nucleus and cluster at corresponding shape deformations $\xi_1$ and $\xi_2$. The value of $\delta E_{12}(R,\xi_1,\xi_2)$ at $R \approx 0$ equals the shell correction of the parent nucleus of the same shape.

The nuclei strongly interact at small distances. This interaction leads to the shift and splitting of the single-particle levels in both nuclei. Due to this, the proton and neutron single-particle spectra around the Fermi levels became more homogenous near the touching distance $R_{t}(\xi_1,\xi_2)$ of nuclei as well as at smaller distances. Such behavior of single-particle levels is clearly demonstrated in the framework of two-center shell model \cite{msd,mg}.

According to the shell-correction prescription \cite{s}, the absolute value of the shell-correction is reduced in the case of more homogenous single-particle spectra around the Fermi levels. The sharp reduction of the shell-correction contribution into the total potential energy around the touching point of cluster and residue nucleus is obtained in the fission approximation of cluster decay in Ref. \cite{msd}. So, the energy-level splitting, which is proportional to the strength of the mutual nucleus-nucleus perturbation, reduces the shell correction.

The perturbation of the single-particle level is enlarged with decreasing distance between surfaces of nuclei and increasing interaction between nucleons belonging to different nuclei. The perturbation potential is related to the density distribution in the nucleus, which induces the disturbance. The density distribution is often parametrized by the Fermi distribution; see also Eqs. (10) and (12). Therefore we approximate the shell-correction for system of interacting nuclei around the touching distance $R_{t}(\xi_1,\xi_2)$ as
\begin{eqnarray}
\delta E_{12}(R,\xi_1,\xi_2)\approx [\delta E_1(\xi_1) + \delta E_2 (\xi_2)] f_{sh}(R,\xi_1,\xi_2).
\end{eqnarray}
Here
\begin{eqnarray}
f_{sh}(R,\xi_1,\xi_2) = 1/\left\{1 + \exp{[(R_{t}(\xi_1,\xi_2)-R)/d_{sh}}]\right\},\; \\
R_{t}(\xi_1,\xi_2) = r_{sh} A_{1}^{1/3} (1 + 2 \xi_1) + r_{sh} A_{2}^{1/3} (1 + 2 \xi_2), \;\;
\end{eqnarray}
where $d_{sh}$ is the diffuseness related to the attenuation of the shell-correction with reduction of distance $R$, $A_i=Z_i + N_i$ is the number of nucleons in nucleus $i \; (i=1,2)$, $r_{sh}$ is the radius parameter. This is a rough approximation, but it can greatly simplify the calculations of the shell-correction around the touching point. Note that the exponential reduction of the shell-correction values related to washing out the shell non-homogeneity of single-particle spectra is often considered in nuclear physics \cite{dh,ms,ist}.

The nose-to-nose orientation of prolate nuclei leads to the lowest value of the potential energy barrier height between nuclei \cite{dp,dp2007}. The cluster emission is a slow process related to barrier penetration; therefore, the nose-to-nose orientation of the fragments at cluster decay is taken into account upon evaluation of the touching distance in Eq. (20). Note that fission fragments at the scission point are oriented nose to nose too.

The shell correction of interacting nuclei, $\delta E_{12}(R,\xi_1,\xi_2)$, is smoothly approaching to the limit of non-interacting nuclei (17) at large distances $R$ between nuclei.

\subsection{Cluster decay half-life}

The cluster-decay half-life $T_{1/2}$ is calculated as
\begin{eqnarray}
T_{1/2} = \ln(2)/[\nu \; S \; T(Q)],
\end{eqnarray}
where $\nu$ is the frequency of assaults of the cluster on the barrier, $S$ is the spectroscopic (or preformation) factor, $T(Q)$ is the transmission coefficient, which shows the probability of penetration through the barrier, and $Q$ is the energy released when the cluster decays.

The frequency of assaults of the cluster on the barrier is
\begin{eqnarray}
\nu=\frac{v_2}{R_0}= \frac{\sqrt{2K_2/(M_N A_2)}}{R_0} ,
\end{eqnarray}
where $v_2$ is the velocity of the cluster in the parent nucleus with radius $R_0=r_0(A_1+A_2)^{1/3}$, $r_0=1.15$ fm, $K_2= M_N A_2 v_2^2/2 =[A_1/(A_1+A_2)]Q$ is the kinetic energy of the cluster and $M_N$ is the mass on nucleon. Correspondingly, $K_1=Q-K_2$ is the recoil energy of the residue nucleus. We choose the spectroscopic factor $S$ equals 1. The cluster is ready for emission according to the density-distribution ansatz (9); therefore, value $S=1$ is reasonable for our approach. Note that the value $S=1$ is also used in the fission approximation to cluster decay.

The transmission coefficient can be obtained in the semiclassical WKB approximation
\begin{eqnarray}
T(Q)=1/\left[1 + \exp\left(2 {\cal A} \right) \right],
\end{eqnarray}
where
\begin{eqnarray}
{\cal A} = \frac{1}{\hbar}
\int_{R_a}^{R_b} dR \sqrt{
2 {\cal B}(R) \left( V_{\rm tot}(R,\xi_1,\xi_2)-Q \right) }
\end{eqnarray}
is the action along the trajectory of cluster emission in multidimensional space $\{R,\xi_1,\xi_2\}$. Here $V_{\rm tot}(R,\xi_1,\xi_2)$ is the total potential energy determined by Eq. (16), $R_a,\xi_{1a},\xi_{2a}$ and $R_b,\xi_{1b},\xi_{2b}$ are the coordinates of inner and outer turning points determined by the equations
$V_{\rm tot}(R_{a(b)},\xi_{1,a(b)},\xi_{2,a(b)})=Q$,
\begin{eqnarray}
{\cal B}(R)=
B_{RR}
+ \sum_{k=1,2} \frac{\partial \xi_k}{\partial R} \left[B_{R\xi_k}
+ \sum_{k^\prime=1,2} B_{\xi_k\xi_{k^\prime}} \frac{\partial \xi_{k^\prime}}{\partial R} \right], \;\;
\end{eqnarray}
is the full inertia, $B_{RR}$, $B_{R\xi_k}$ and $B_{\xi_k\xi_{k^\prime}}$ are the mass parameters related to the corresponding collective coordinates. Expressions for accurate evaluation of the mass parameters are given in Ref. \cite{s}.

The nondiagonal terms of mass tensor $B_{ij}$ are zero at large distances between nuclei and negligible at distances slightly larger than the touching distances $R_t(\xi_1,\xi_2)$. Therefore the full inertia at distances $R \geq R_t(\xi_1,\xi_2)$ is
\begin{eqnarray}
{\cal B}(R)=
B_{RR}
+ \sum_{k=1,2} B_{\xi_k \xi_k} \left(\frac{\partial \xi_{k}}{\partial R} \right)^2,
\end{eqnarray}
where
\begin{eqnarray}
B_{RR} = M_N A_1 A_2/(A_1+A_2), \\
B_{\xi_1\xi_1} = k_1 B_{1\; {\rm irrot}} = k_1 \frac{6}{5} M_N r_0^2 A_1^{5/3} ,\\
B_{\xi_2\xi_2} = k_2 B_{2\; {\rm irrot}} = k_2 \frac{6}{5} M_N r_0^2 A_2^{5/3}.
\end{eqnarray}
We use hydrodynamical mass parameter for irrotational flow \cite{bohrmott} for mass parameters $B_{i\; {\rm irrot}}$ [see Eqs. (28)-(29)], which are modified due to coupling between deformation parameters $\xi$ and $\beta_2$. It is well known, that the value of hydrodynamical mass parameter for irrotational flow is much smaller than the realistic one \cite{pg,s,kp,sob}, therefore, we introduce the enhancement factor $k_i$ in Eqs. (28) and (29). This is a rough approximation, which is reasonable for small values of deformations, when a detailed dependence of $B_{\xi_i\xi_i}$ on the deformation value $\xi_i$ is not important. The values of the enhancement coefficient $k_i$ can be evaluated from the ratio
\begin{eqnarray}
k_i=\frac{B_{i\; {\rm ho}}}{B_{i\; {\rm irrot}}} = \frac{15 \hbar^2 Z_i^2 e^2 r_0^2 }{4 M_N E_{2i} A_i^{1/3} \pi B_i(E2,0\rightarrow 2)},
\end{eqnarray}
where $B_{i\; {\rm ho}}$ is the mass parameter of harmonic quadrupole surface oscillations with energy $E_{2i}$ and $B_i(E2,0 \rightarrow 2)$ is the value of the reduced transition probability in nucleus $i$ \cite{bohrmott}. The experimental values of $E_{2i}$ and $B_i(E2,0 \rightarrow 2)$ can be found in Ref. \cite{be2}.

At distances $R \leq R_t(\xi_1,\xi_2)$ we approximate the full inertia as
\begin{eqnarray}
{\cal B}(R) \approx {\cal B}(R_t(\xi_1,\xi_2)) \; k_0^{\left[ 1- R/R_t(\xi_1,\xi_2)\right]} ,
\end{eqnarray}
where ${\cal B}(R_t(\xi_1,\xi_2))$ is the full inertia at the touching point described by Eq. (26) and $k_0$ is the enhancement factor. Similar inertia parametrizations are often used in phenomenological approaches to fission \cite{kp,sob}.

\section{Results and discussion}

The multidimensional model of cluster decay combines the basic idea of fission and cluster preformation approaches, because the potential energy (16) and the action (24) are evaluated in a similar way as those in some fission models \cite{spg,pg,hrm,kp,s} and the cluster exists in the parent nucleus as proposed in the cluster-preformation models.

At the beginning we consider the macroscopic potential energy $V_{\rm macro}(R,\xi_1,\xi_2)$ described by Eqs. (1)--(15) for cluster decay $^{242}$Cm $\rightarrow \; ^{208}$Pb + $^{34}$Si. This potential energy surface is presented in Fig. 1 for the case $\xi_1=\xi_2=\xi$. We evaluate $V_{\rm macro}(R,\xi,\xi)$ for the parameter set SkM$^{\star}$ of the Skyrme force \cite{skmstar}, because the fission properties of actinides are accurately described by using this set \cite{bgh,skmstar}. Moreover, the heights of nucleus-nucleus potential barriers for various systems, which are also important for cluster decay, evaluated for this parameter set of Skyrme force agree well with the empirical ones \cite{d_barr,duttp}.

\begin{figure}
\begin{center}
\includegraphics[width=8.7cm]{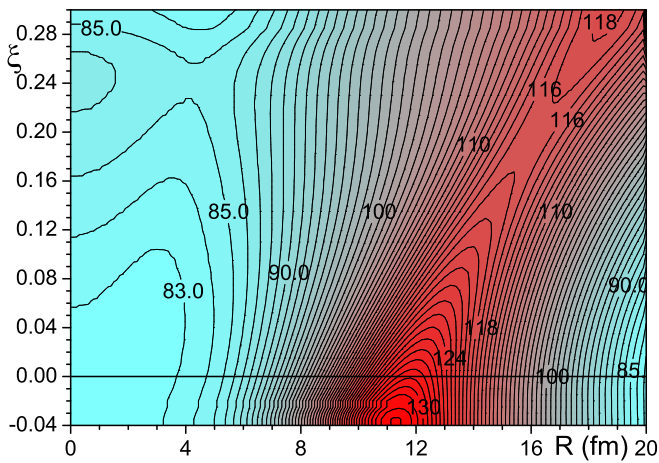}
\caption{(Color online) The macroscopic potential energy $V_{\rm macro}(R,\xi,\xi)$ for cluster decay $^{242}$Cm $\rightarrow \; ^{208}$Pb + $^{34}$Si.}
\end{center}
\end{figure}

The potential energy surface $V_{\rm macro}(R,\xi,\xi)$ (see Fig. 1) is very flat in the range $0\leq R \lesssim R_{\rm Pb}-R_{\rm Si} \approx 3 \; {\rm fm}$ at small $\xi$. Here $R_{\rm Pb}\approx6.8 \; {\rm fm}$ is the radius of $^{208}$Pb and $R_{\rm Si}\approx 3.7 \; {\rm fm}$ is the radius of $^{34}$Si. The cluster is located inside the volume of residue nucleus at such distances and, due to this, the shape of the parent nucleus is not disturbed. Therefore, a flat shape of the potential surface at small $R$ is natural and the cluster can easily move inside the parent nucleus.

At larger distances $R \gtrsim 4$ fm the cluster starts to form the bump on the surface of the parent nucleus and therefore the potential energy starts to rise. The growth of potential surface is strong around the touching distances $R_{t}(\xi_1,\xi_2)$ of the residue nucleus and cluster. Note that $R_{t}(0,0)=R_{\rm Pb}+R_{\rm Si} \approx 10.5$ fm for case of spherical nuclei. The potential surface rises until the barrier distance $R_{b}(\xi_1,\xi_2)$ ($R_{b}(\xi_1,\xi_2)> R_{t}(\xi_1,\xi_2)$) and smoothly decreases at distances beyond the barrier. The ridge, which we can see in Fig. 1, separates deformed one-body shapes (or closely spaced two-body shapes) and strongly separated two-body shapes.

The dependence of the macroscopic potential energy along axis $\xi$ at small $R$ is related to the macroscopic fission barrier of the parent nucleus $^{242}$Cm induced by large ellipsoidal deformation. The macroscopic fission barrier takes place at $\xi \approx 0.24$ and $R \approx 4$ fm, see Fig. 1. The barrier height relatively to the ground-state energy of the parent nucleus is close to 2.5 MeV. This value of the macroscopic fission barrier is very close to the 2.75 MeV evaluated in Ref. \cite{sierk} in another macroscopic approach. The fission barrier height is very low in comparison to the height of the ridge, related to the cluster decay $^{242}$Cm $\rightarrow \; ^{208}$Pb + $^{34}$Si, see Fig. 1. For example, the cluster decay barrier height relatively to the ground-state energy of the parent nucleus is close to 47 MeV for spherical nuclei.

The height of the ridge separated one-body and two-body forms is reduced with increasing $\xi$, see Fig. 1. Therefore a cluster emission trajectories, which pass via points with $\xi_{1,2}>0$, may lead to smaller values of action (24), as the result, transmission coefficient (23) can be drastically enhanced due to exponential dependence on the action. As pointed out in the introduction, a similar effect is very important for subbarrier fusion of heavy ions.

The lowest value of the potential at large distances between nuclei $R$ takes place for slightly oblate ($\xi<0$) shapes, as discussed in Ref. \cite{d90} and papers cited therein.

\begin{figure}
\begin{center}
\includegraphics[width=8.7cm]{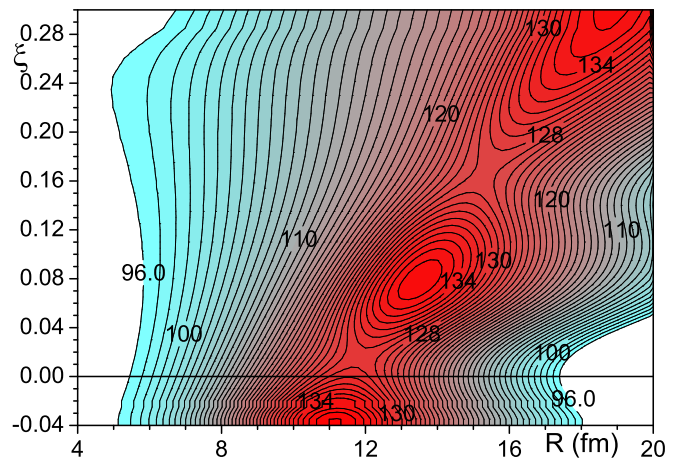}
\caption{(Color online) The total (macroscopic $+$ microscopic) potential energy $V_{\rm tot}(R,\xi,\xi)$ for cluster decay $^{242}$Cm $\rightarrow \; ^{208}$Pb + $^{34}$Si.}
\end{center}
\end{figure}

Now we add the shell-corrections (18)--(20) to the macroscopic part of the potential energy $V_{\rm macro}(R,\xi_1,\xi_2)$ and find the total potential energy $V_{\rm tot}(R,\xi_1,\xi_2)$ (16). We evaluate the proton and neutron single-particle levels in $^{208}$Pb and $^{34}$Si by using WSBETA code for the universal parameter set of the Woods-Saxon proton and neutron mean fields \cite{wsbeta}. Using these levels we find the shell corrections for the residue nucleus $\delta E_1(\xi_1)$ and the cluster $\delta E_2 (\xi_2)$ for the standard parameter values of the shell-correction prescription \cite{s}.

The total (macroscopic-microscopic) potential energy surface $V_{\rm tot}(R,\xi_1,\xi_2)$ (16) for cluster decay $^{242}$Cm $\rightarrow \; ^{208}$Pb + $^{34}$Si is presented in Fig. 2 for case $\xi_1=\xi_2=\xi$ and $r_{sh}=1.15$ fm. The value of $d_{sh}$ [see Eq. (19)] is related to the diffuseness of density distribution in nucleus induced the perturbations of single-particle levels and the range of nucleon-nucleon force. The typical values of diffuseness of density distribution in nuclei are close to 0.5$\div$0.55 fm \cite{vjv}. Taking into account the finite range of the nucleon-nucleon force we choose $d_{sh}=0.6$ fm. Our approach for shell-correction (18)-(20) is reliable for large distances $R$ as well as for distances around the touching points; therefore, we cut the map for small $R$. Note that the experimental value of the energy released upon the cluster decay $^{242}$Cm $\rightarrow \; ^{208}$Pb + $^{34}$Si is $Q=96.5$ MeV \cite{ame2012}, therefore, the contour lines related to 96 MeV correspond to the lowest value of energy on the potential energy surface pointed out in Fig. 2.

Comparing the landscapes of macroscopic $V_{\rm macro}(R,\xi,\xi)$ and total $V_{\rm tot}(R,\xi,\xi)$ potential energies in Figs. 1 and 2 we see drastic changes induced by the contribution of the shell-correction of $^{208}$Pb into the potential energy surface in Fig. 2.

The values of macroscopic-microscopic and macroscopic potential energies in Figs. 1 and 2 are significantly different at the distances around the touching points and barriers as for spherical as for deformed nuclei. The shell correction contribution to the total potential energy between $^{208}$Pb and $^{34}$Si enlarges the height of the barrier to $\approx 1$ MeV for the case of both spherical nuclei. Moreover, the difference between the potentials $V_{\rm tot}(R,\xi_1,\xi_2)$ and $V_{\rm macro}(R,\xi_1,\xi_2)$ increases at smaller values of $R$ because of the radial dependence of the shell correction contribution [see Eqs. (18)-(20)].

There are two dips in the ridge on the way for cluster emission from small to large values $R$ in Fig. 2. The dip at $\xi \sim 0.02$ is related to slightly deformed (near spherical) shapes of interacting nuclei, while the dip at $\xi \sim 0.16$ is linked to strongly elongated shapes of nuclei. These dips are related to the deformation dependence of the shell correction value $\delta E(\xi)$. The trajectories passing through these dips may have the lowest values of action.

We parametrize the dependence of the deformation of residue ($i = 1$) and cluster ($i = 2$) nuclei on $R$ along the path of the cluster emission by polynomial
\begin{eqnarray}
\xi_i(R)&=&a_{i1} s+a_{i2} s^2+a_{i3} s^3 + \xi_i^0 ,
\end{eqnarray}
where $s=R/R_b-1$, $R_b$ is the coordinate of the outer turning point, $\xi_i^0$ are the ground-state deformation of nuclei and $a_{ij}$ are 6 variational parameters, which are found by the numerical minimization of the action (24). The cluster decay into the spherical residue and cluster nuclei leads to $\xi_i(R_b)=\xi_1^0=\xi_2^0=0$ for $i=1,2$. In contrast to this, the shapes of cluster and residue nucleus are not fixed at the inner turning point $R_a$ in the framework of our multidimensional cluster preformation model.

The values of the mass parameter enhancement factor $k_i$ [see Eq. (30)], are evaluated using the experimental values \cite{be2} of the energies $E_{2i}$ and the reduced transition probability $B_i(E2,0 \rightarrow 2)$ of harmonic quadrupole surface oscillations in nuclei $^{208}$Pb and $^{34}$Si. The values of these factors are $k_1=6.07$ and $k_2=14.01$, respectively.

The half-lives of cluster decay $^{242}$Cm $ \rightarrow \; ^{208}$Pb + $^{34}$Si obtained in our model for various variants of shape deformations of nuclei are presented in Table 1. The experimental value of half-life for this cluster decay is $1.4^{+0.5}_{-0.3} \cdot 10^{23}$ s \cite{ogloblin2000}. This value is well described in our model, when we take into account dynamical deformation in both nuclei and realistic values of mass parameter enhancement factors $k_0=3.35$, $k_1=6.07$ and $k_2=14.01$; see Table 1.

\begin{table}
\caption{The values of cluster decay half-lives obtained for various types of trajectories. The values of cluster decay half-lives are evaluated for trajectories related to both spherical nuclei $T_{\rm sph}$, spherical cluster and dynamically deformed lead $T_{\rm Pb}$, dynamically deformed cluster and spherical lead $T_{\rm Si}$ and both dynamically deformed nuclei $T_{\rm Pb+Si}$.}
\begin{tabular}{ccccccc}
\hline \hline
$k_0$ &$k_1$ &$k_2$ & $T_{\rm sph}$ (s) & $T_{\rm Pb}$ (s) & $T_{\rm Si}$ (s) & $T_{\rm Pb+Si}$ (s) \\
\hline
1 & 1 & 1 & $2.5 \cdot 10^{24}$ & $4.9 \cdot 10^{18}$ & $9.4 \cdot 10^{23}$&
$3.9 \cdot 10^{18}$\\
3.35 & 1 & 14.01 & $2.2 \cdot 10^{26}$ & $6.9 \cdot 10^{19}$ & $7.1 \cdot 10^{25}$&
$4.6 \cdot 10^{19}$\\
3.35 & 6.07 & 1 & $2.2 \cdot 10^{26}$ & $1.8 \cdot 10^{23}$ & $6.6 \cdot 10^{25}$&
$1.3 \cdot 10^{23}$\\
3.35 & 6.07 & 14.01 & $2.2 \cdot 10^{26}$ & $1.8 \cdot 10^{23}$ & $7.1 \cdot 10^{25}$&
$1.4 \cdot 10^{23}$\\
\hline \hline
\end{tabular}
\end{table}

As example, we evaluate the cluster decay half-lives for the case $k_0=k_1=k_2=1$ and various variants of surface deformations of nuclei. These evaluations are related to irrotational hydrodynamic flow of nucleons induced by surface deformation of nuclei. The values of the cluster-decay half-lives are lowest for such cases; see Table 1. The value of half-life rises with increasing $k_0$.

We remind the reader that the shell-correction contribution to the interaction potential is taken into account in both the fission theory and the fission approach for cluster decay. In contrast to this, the shell-correction contribution to the interaction potential energy of nuclei is ignored in the cluster-preformation approach.

The cluster-decay half-life evaluated for the spherical nuclei emission path and without the shell-correction contribution to the cluster-nucleus interaction potential energy (16) for the case $k_0=k_1=k_2=1$ is $T_{\rm sph}^{\rm without\; shell \; corr}=8.3 \cdot 10^{15}$ s. This value is much less than $T_{\rm sph}=2.5 \cdot 10^{24}$ s evaluated with the shell-correction contribution for the corresponding case; see Table 1. We emphasize that the values of macroscopic-microscopic interaction potential energy around the touching point are larger than the values of the macroscopic one due to large negative shell correction value in lead [see Eqs. (16), (18)--(20)], therefore $T_{\rm sph}^{\rm without\; shell \; corr} << T_{\rm sph}$.

Note that the correct description of cluster-decay half-life in the framework of the cluster preformation model without the shell-correction contribution to the cluster-nucleus interaction potential can be obtained by introduction of the spectroscopic factor (or cluster preformation probability). Spectroscopic factors are often used in various cluster preformation models (see, for example, Refs. \cite{bw,k} and Refs. cited therein).

\begin{figure}
\begin{center}
\includegraphics[width=8.7cm]{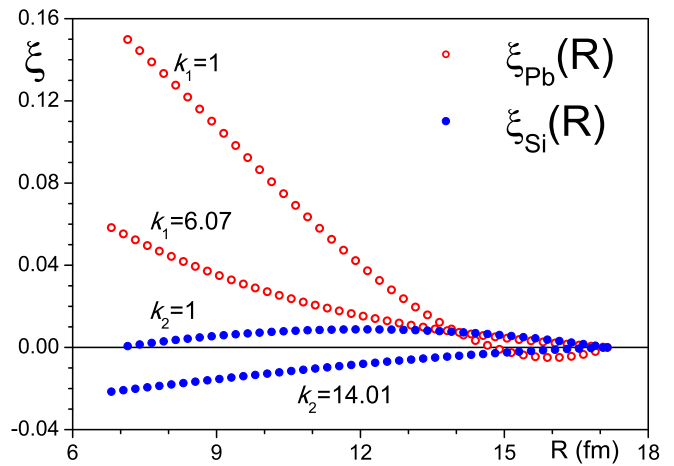}
\caption{(Color online) The dependencies of deformation parameters for the cluster $\xi_{\rm Si}(R)$ and residue nucleus $\xi_{\rm Pb}(R)$ evaluated at hydrodynamical irrotational and realistic values of mass parameter enhancement factor $k_i$ for cluster decay $^{242}$Cm $\rightarrow \; ^{208}$Pb + $^{34}$Si.}
\end{center}
\end{figure}

The dependencies of deformation parameters for the cluster $\xi_{\rm Si}(R)$ and residue nucleus $\xi_{\rm Pb}(R)$ on the distance $R$ are presented for hydrodynamic irrotational (at $k_1=k_2=1$) and realistic (at $k_1=6.07, \; k_2=14.01$) values of the mass parameters in Fig. 3.

The accuracy of enhancement coefficients $k_1$ and $k_2$ is related to the accuracy of $B_i(E2,0 \rightarrow 2)$ evaluation; see Eq. (30). The experimental errors of $B_i(E2,0 \rightarrow 2)$ are given in Ref. \cite{be2}; therefore, accuracies of coefficients $k_1$ and $k_2$ are close to 10\% and 40\%, respectively. If we change the value of $k_0$ or $k_1$ by 10\%, then the cluster decay half-life varies  $\approx$40\% or $\approx$35\%, respectively. Thus the cluster decay half-life strongly depends on values of enhancement coefficients $k_0$ and $k_1$. In contrast to this, if we shift the value of $k_2$ by 40\%, then the cluster-decay half-life changes by $\approx$20\%. The value of the enhancement coefficient of cluster $k_2$ weakly affects the cluster decay half-life.

The consideration of cluster decays $^{228}$Th $\rightarrow \, ^{208}$Pb + $^{20}$O and $^{242}$Cm $\rightarrow \, ^{208}$Pb + $^{34}$Si is similar in the framework of the multidimensional model, because the ground-state shapes of nuclei $^{208}$Pb, $^{20}$O, and $^{34}$Si are spherical. Using experimental data \cite{be2} for the energy and the reduced transition probability of harmonic quadrupole surface oscillations in nucleus $^{20}$O, we find the enhancement coefficient $k_2=32.85$. The other enhancement coefficients $k_0$ and $k_1$ are the same as before. The half-life of cluster decay $^{228}$Th $\rightarrow \, ^{208}$Pb + $^{20}$O obtained in our model is $5.0 \cdot 10^{20}$ s. This value is very close to the experimental value $(5.29 \pm 1.01) \cdot 10^{20}$ s \cite{bonetti1993}.

Applying our consideration to decays $^{232}$U $\rightarrow \, ^{208}$Pb + $^{24}$Ne and $^{236}$Pu $\rightarrow \, ^{208}$Pb + $^{28}$Mg and substituting into Eq. (32) the values of the ground-state quadrupole deformation of the surface of $^{24}$Ne and $^{28}$Mg from Ref. \cite{mnms}, we evaluate the half-lives for these decays in the framework of the multidimensional model, which are $5.8 \cdot 10^{19}$ s and $1.0 \cdot 10^{21}$ s, respectively. The values of enhancement coefficients $k_0$ and $k_1$ at half-life calculations are the same as before, but the values of $k_2$ are obtained by using experimental data for the energy and the reduced transition probability of harmonic quadrupole surface oscillations in $^{24}$Ne and $^{28}$Mg, see Ref. \cite{be2}. The theoretical values of half-lives are three to four times smaller than the corresponding experimental values $(2.4 \pm 0.2) \cdot 10^{20}$ s \cite{bonetti1991} and $3.3^{+0.7}_{-1.2} \cdot 10^{21}$ s \cite{hussonnois1995}. Note that the cluster decay half-lives are very large; therefore, description of the experimental data with the accuracy of one or two orders of magnitude is considered  good; see, for example, results for cluster-decay half-lives in the recent approach to cluster decay in Ref. \cite{zzyq}. Therefore, our description of half-lives for decays $^{232}$U $\rightarrow \, ^{208}$Pb + $^{24}$Ne and $^{236}$Pu $\rightarrow \, ^{208}$Pb + $^{28}$Mg is very good.

\begin{figure}
\begin{center}
\includegraphics[width=8.7cm]{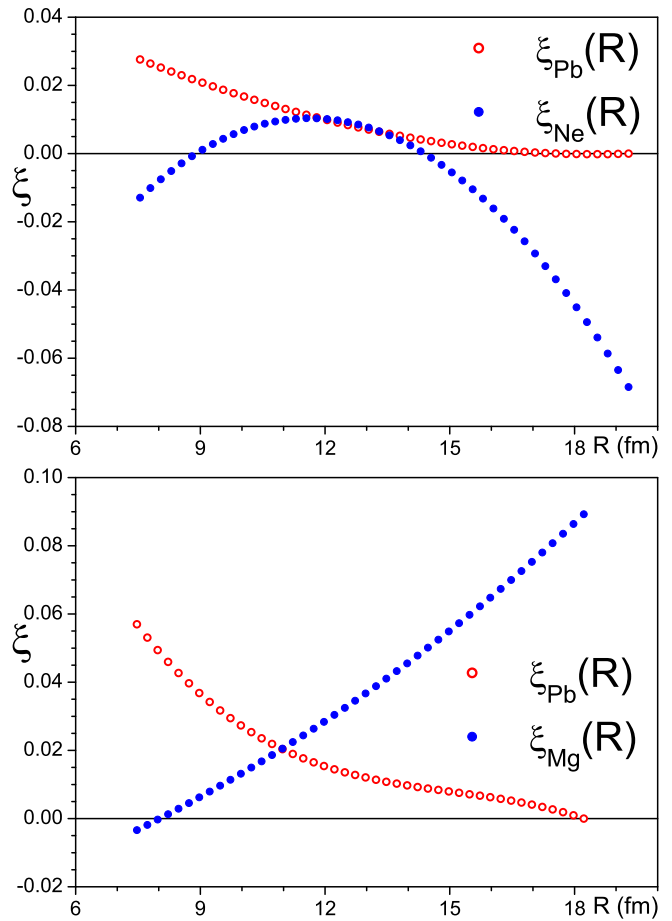}
\caption{(Color online) The dependencies of deformation parameters for the cluster and residue nucleus evaluated at realistic values of mass parameter enhancement factor $k_i$ for cluster decays $^{232}$U $\rightarrow \, ^{208}$Pb + $^{24}$Ne (upper panel) and $^{236}$Pu $\rightarrow \, ^{208}$Pb + $^{28}$Mg (bottom panel).}
\end{center}
\end{figure}

The dependencies of deformation parameters for the cluster and residue nucleus on the distances $R$ are presented for realistic values of the mass parameters for cluster decays $^{232}$U $\rightarrow \, ^{208}$Pb + $^{24}$Ne and $^{236}$Pu $\rightarrow \, ^{208}$Pb + $^{28}$Mg in Fig. 4. Nucleus $^{24}$Ne has oblate ground-state surface deformation \cite{mnms}; therefore, oblate surface deformation of $^{24}$Ne leads to a rise of absolute values and tends to the ground-state value during cluster decay. However the deformation of $^{24}$Ne is prolate around barrier distances. In contrast to this nucleus $^{28}$Mg has prolate ground-state surface deformation \cite{mnms}; therefore, prolate surface deformation of $^{28}$Mg tends to the ground-state value during cluster emission. Deformation of $^{208}$Pb is decreased from slightly prolate to spherical ground-state value. The dependencies of surface deformation of lead on $R$ are similar for various cluster-emission cases, see Figs. 3 and 4.

Analyzing results presented here we make the following conclusions:
\begin{itemize}

\item[--] The paths of cluster decay at $k_i=1$ are related to very deformed shapes at small distances $R$. These trajectories lead to the lowest half-life value. These results are related to unrealistically small values of mass parameters $B_{\xi_i\xi_i}$.

\item[--] The trajectories of cluster decay at realistic values of $k_i$ pass through the slightly deformed shapes. The shape of residue nucleus is prolate along the cluster decay path.

\item[--] The dynamical deformation of the cluster nucleus depends on the ground-state deformation of the cluster strongly.

\item[--] The dynamical deformation of the residue nucleus effects the decay half-life much stronger than the dynamical deformation of the cluster.

\item[--] The values of half-life evaluated with dynamical deformation are much smaller than the one without dynamical deformation.

\item[--] The shell-correction contribution to total interaction potential between the cluster and the residue nucleus is very important for the potential energy landscape as well as for the half-life evaluation.

\end{itemize}

\section{Conclusion}
The half-lives of cluster decays $^{228}$Th $\rightarrow \, ^{208}$Pb + $^{20}$O, $^{232}$U $\rightarrow \, ^{208}$Pb + $^{24}$Ne, $^{236}$Pu $\rightarrow \, ^{208}$Pb + $^{28}$Mg and $^{242}$Cm $ \rightarrow \; ^{208}$Pb + $^{34}$Si are successfully described in the framework the multidimensional cluster preformation model. Only one fitting parameter $k_0$ is fixed for cluster decay $^{242}$Cm $ \rightarrow \; ^{208}$Pb + $^{34}$Si. The half-lives of cluster decay for other cases are evaluated by using this value of $k_0$.

The macroscopic energy of interacting nuclei in this model is evaluated in the framework of nonlocal $\hbar^4$ ETF approach with the Skyrme and Coulomb forces. The relaxed proton and neutron densities of cluster and residue nuclei described by ansatz (9) are used along the cluster-decay path.

The shell correction contribution to the total nucleus-nucleus potential is very important around the touching point and barrier ridge. The large negative value of shell-correction in the ground-state of $^{208}$Pb drastically changes both the potential energy landscape and the cluster-decay half-life.

The heights of the barrier between prolate nuclei are lower than the height of the barrier between spherical nuclei. Therefore the dynamical deformations of nuclei increase the barrier penetrability and reduce the half-life value. The influence of dynamical deformations of nuclear shape around the barrier is very important for an accurate description of cluster emission half-life.

The ground-state deformation of cluster strongly influences the trajectory of cluster emission during cluster decay.

The shape of trajectories and the cluster decay half-life depend strongly on the mass parameters values (or the mass parameter enhancement factors $k_i$).

\section*{Acknowledgements}

The author thanks Dr. V. I. Tretyak for useful remarks.

\end{document}